\documentclass[reprint,jap,aip]{revtex4-1}
\usepackage{amssymb}
\usepackage{amsmath}
\usepackage{graphicx}
\usepackage{dcolumn}
\usepackage{bm}
\usepackage[latin1]{inputenc}

\usepackage[active]{srcltx}
\usepackage{ae,aecompl}

\usepackage{pslatex}
\usepackage[T1]{fontenc}
\AtBeginDocument{\RequirePackage{lmodern, times}}

\usepackage{color}
\newcommand{\Phineg}{\boldsymbol{\Phi}}
\newcommand{\Psineg}{\boldsymbol{\Psi}}

\newcommand{\bra}{\langle}
\newcommand{\ket}{\rangle}
\begin{document}


\title[Quantum transport]{Interference effects  induced by Andreev bound states in a hybrid nanostructure composed by a quantum dot coupled to ferromagnetic and superconductor leads}
\author{E. C. Siqueira$^{*1,2}$, P. A. Orellana$^{3}$, A. C. Seridonio$^{2}$, R. C. Cestari$^{2}$, M. S. Figueira$^{4}$, G. G. Cabrera$^{5}$}
\affiliation{$^{1}$ Departamento de Física, Universidade Tecnológica Federal do Paraná (UTFPR), 84016-210, Ponta Grossa, PR, Brasil\email{ezcostta@gmail.com}\\
$^{2}$Departamento de Física e Química, Universidade Estadual Paulista (UNESP),  15385-000, Ilha Solteira, SP, Brasil\\
$^{3}$ Departamento de Física, Universidad Técnica Federico Santa Maria, Av. Vicuña Mackenna 3939, Santiago, Chile\\
$^{4}$Instituto de F\'{i}sica, Universidade Federal Fluminense, 24210-340, Niter\'oi, RJ, Brasil\\
$^{5}$Instituto de Física  `Gleb Wataghin', Universidade Estadual de Campinas (UNICAMP), Campinas 13083-859, SP, Brasil}

\begin{abstract}
In this work, it is considered a nanostructure composed by a quantum dot  coupled to two ferromagnets and a superconductor. The transport properties of this system are studied within a generalized mean-field approximation taking into account proximity effects and spin-flip correlations within the quantum dot.   It is shown that the zero-bias transmittance for the co-tunneling between the ferromagnetic leads presents a dip whose height depends on the relative orientation of the magnetizations. When the superconductor is coupled to the system, electron-hole correlations between different spin states leads to a resonance in the place of the dip appearing in the transmittance. Such an effect is accompanied by two anti-resonances explained by a ``leakage''~of conduction channels from the co-tunneling to the Andreev transport. In the non-equilibrium regime, correlations within the quantum dot introduce a dependence of the resonance condition on the finite bias applied to the ferromagnetic leads. However, it is still possible to observe signatures of the same interference effect in the electrical current.
\end{abstract}


\maketitle





\section{Introduction}

Conventional superconductivity (\emph{s}-wave) and ferromagnetism present different spin symmetries required by their order parameters.  In \emph{s}-wave superconductors, the Cooper pairs are in the singlet state while exchange interaction induces a triplet alignment in ferromagnets. As a result, superconductivity is strongly suppressed or even completely destroyed in bulk compounds in which both order parameters would be present \cite{Izyumov}. On the other hand, the production of superconductor/ferromagnet (F/S) layered systems has allowed the study of the interplay between superconductivity and ferromagnetism \cite{Buzdin} experimentally. In these systems, superconductor and ferromagnet are spatially separated by a narrow interface whereby Cooper pairs can diffuse into the ferromagnet. Conversely, spin polarization can be induced into the superconductor by diffusion of ferromagnetic electrons. These effects are called \emph{proximity effects} and are the responsible for particular features of F/S systems. In fact, the thermal, magnetic and electrical properties of these systems are completely different  in comparison to the ferromagnet and superconductors in their bulk form \cite{Buzdin,franceschi}. Within the vast set of phenomena, one can highlight the $\pi$-phase transition of the superconductor order parameter in  S/F/S~Josephson junctions \cite{buzdin22}, the nonmonotonic behavior of the superconducting critical temperature $(T_{c})$ on the layers thicknesses in multilayers systems \cite{Jiang314} and oscillations on electronic density of states \cite{Buzdin}.
\begin{figure}[h]\centering
\includegraphics[scale=1]{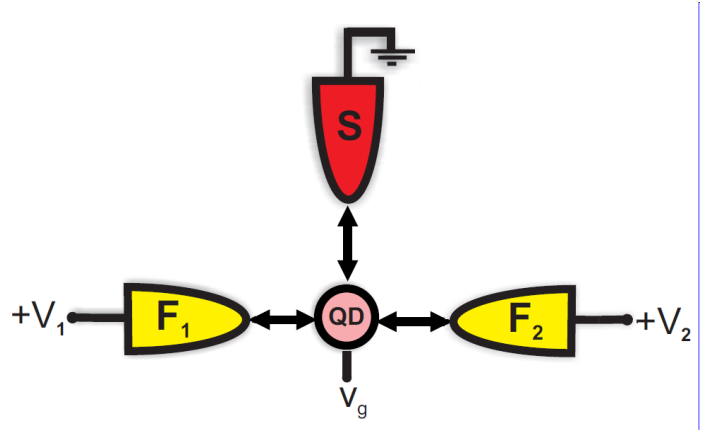}
\caption{\label{fig1}(Color Online)~Schematic diagram for the
\textit{F}$_{1}$-(\textit{QD}-\textit{S})-\textit{F}$_{2}$ system. The magnetization of $F_{1}$ is assumed to be fixed and the magnetization of $F_{2}$ can be varied of an angle $\theta$ with respect to the $F_{1}$ magnetization. $V_{1}$ and $V_{2}$ are the external potentials applied to  $F_{1}$ and $F_{2}$, respectively, while the superconductor is grounded. $V_{g}$ being the gate potential applied to the QD.}
\end{figure}

Concerning transport properties, F/S systems yield controlling the charge current through the spin degree of freedom which is of interest in areas like spintronics\cite{Eschrig,fabianJ}. In fact,  for bias voltages within the superconductor gap, the current is carried via Andreev reflections (ARs)\cite{andreevref,Beenakker1}. In this process, an incident electron of a given energy $E$ and spin up (down)  is reflected on the superconductor as a hole of energy $-E$ and spin down (up). As a result, the current in the ferromagnet  is converted into a Cooper pair current in the superconductor. The current is different of zero only if there are available states for both spins around the Fermi level. However, the occupation of these states is dependent on the ferromagnet polarization. When the ferromagnet polarization ($P$) is equal to unity  there are no available states for holes with spin down and the system behaves as an insulator; for $P=0$ the ferromagnet is unpolarized and the current reaches a maximum value.

The electrons forming the Cooper pair are highly correlated in large distances in comparison to interatomic distances. This feature has been explored by Deutscher and Feinberg \cite{deutscher} to propose a non-local Andreev reflection (called crossed AR), where two electrons of different leads can combine into a Cooper pair if the distance between these leads is smaller than the superconductor coherence length\cite{BCS,btk}. Since this proposal, there has been a profusion of works exploring crossed AR  in different conduction regimes \cite{Brinkman,Lambert4} (ballistic and diffusive) and other materials like superconductor-induced graphene\cite{BeenakkerGraph} and cuprate superconductors\cite{AndreevSaintJames}.


With the recent technology of production of quantum dots (QDs), it is also possible to implement hybrid nanoelectronic devices combining superconductivity and ferromagnetism. In these systems one is able to investigate purely quantum phenomena and the conduction of electrical current with a discrete flux of charges through the QDs\cite{Gustavsson}. As a result, new phenomena may arise concerning the transport properties of these systems\cite{spininjector,Xiufeng,Feng,Sun,Dolcini,Songintra, Circuit,Benjamin,Peysson,Giazotto,DeutscherCross,Melin,Citro,Bai1,Bai2,longbai}.  A particular feature of these hybrid F/S systems
is the current dependence on QD spectral properties, besides the ferromagnet polarization. Additionally, electrons are squeezed into the QD through which the current is injected into the superconductor. As a result, the simplest theoretical treatment for these systems must take into account the Coulomb correlations inside the QDs\cite{Gustavsson,haug}.

In the aforementioned papers, such a correlation is treated by using a perturbative approach since its many-body feature forbids any exact approach to the problem.  Many approximation schemes to treat the electronic correlation exist in the literature in order to better describe the current flow in these hybrid systems. However, as mentioned in the first paragraph, systems involving superconductors exhibit the so-called proximity effect which induces pair correlations in the material in contact with the superconductor\cite{Buzdin}. The interplay between pair correlations with Coulomb interaction may lead to rather complex spectral properties for the QD\cite{bulka}. More specifically, a multi-peak structure has been observed in a $F-QD-S$ nanostructure caused by the interplay between Coulomb correlations and Zeeman splitting due to an external magnetic field\cite{bocian}.

In order to explore the role of the QD spectral properties on the transport of F/S systems, we consider in this work a three-terminal $F_{1}-(QD,S)-F_{2}$ nanostructure as illustrated in Fig. \ref{fig1}.  In this system, the QD is coupled to two ferromagnets in such a way that a voltage bias is applied to $F_{1}$ while $F_{2}$ is grounded. The transmittance with and without the presence of the superconductor lead is considered in order to determine the role of the AR in the co-tunneling current between $F_{1}$ and $F_{2}$. That orientation of the magnetization for the lead $F_{1}$  is fixed while the magnetization of $F_{2}$ is directed to an angle $\theta$ with respect to $F_{1}$. It may be varied from 0 (parallel configuration) to $\pi$~(antiparallel configuration). The correlations within the QD are treated by using a generalized mean-field approximation taking into account proximity effects due to the superconductor and spin-flip processes within the QD. While the spin-flip process has been addressed in a phenomenological way in previous works\cite{spinflip1,spinflip2}, here such an effect is a natural result of the interplay between Coulomb correlations and the misalignment of the magnetizations from the ferromagnetic leads.
It worth mentioning that the geometry shown in Fig. \ref{fig1} has already been studied considering the situation in which the QD is noninteracting and considering the non-local transport due to crossed AR\cite{principal,fengfusheng}. More recently, spin-dependent conductance and thermoelectric properties were addressed for this nanostructure where the role of AR are considered\cite{WeymannThermopower}. In Ref. \onlinecite{Weymannproximity}, the differential conductance and the magnetoresistance have been studied in which a zero-bias anomaly in the Andreev conductance is reported. Such an anomaly is explained by the spin-accumulation generated within the QD due to the coupling to ferromagnets.  In this work, we focus on the co-tunneling process and how it is affected by the coupling to the superconductor.  We observe a resonance appearing in the transmittance due to the interplay between different spin-channels and the Andreev bound states within the QD.

This paper is organized as follows: in Sec. \ref{modelggg} we present the model for the system displayed in Fig. \ref{fig1} and the physical quantities are determined by using the formalism of non-equilibrium Green's functions. In Sec. \ref{resdic} the results are presented and discussed. Finally, a summary and the main conclusions are presented in Sec. \ref{concl}. \\


\section{\label{modelggg}Model and Formulation}

In this section we provide a general description of the formalism to be used to carry out the calculations of the physical quantities. We have used the Keldysh formalism within the Nambu notation\cite{claro,claro2} which allows us to describe spin and electron-hole degrees of freedom in the same footing. This is widely used to tackle systems involving ferromagnets and superconductors.

\subsection{Hamiltonian}

The Hamiltonian is given by a sum of terms describing each part of the system illustrated in Fig. \ref{fig1}. The ferromagnetic and superconductor leads are considered to be non-interacting in such a way that mean field theories can be applied to model these leads. The quantum dot is considered to be composed by a single level spin degenerated with the presence of Coulomb correlations. The coupling between the QD and leads is taken into account phenomenologically by means of a tunneling Hamiltonian.  In this way, the full Hamiltonian is written as
\begin{align}\label{model}
\hat{\mathcal{H}}=\hat{\mathcal{H}}_{1}+\hat{\mathcal{H}}_{2}+\hat{\mathcal{H}}_{S}+\hat{\mathcal{H}}_{\mathcal{C}}+\hat{\mathcal{H}}_{T}.
\end{align}

The terms $\hat{\mathcal{H}}_{1}$ and $\hat{\mathcal{H}}_{2}$ are the Hamiltonians describing the ferromagnets $F_{1}$ and $F_{2}$, respectively. Explicitly, these are given by:
\begin{align}\label{model:F1}
\hat{\mathcal{H}}_{1}=\sum_{k}\hat{\Phineg}^{\dag}_{1k}\hat{\mathbf{E}}_{1,k}(0)\hat{\Phineg}_{1k}
\end{align}
and
\begin{align}\label{model:F2}
\hat{\mathcal{H}}_{2}=\sum_{k}\hat{\Phineg}^{\dag}_{2k}\hat{\mathbf{E}}_{2,k}(\theta)\hat{\Phineg}_{2k}
\end{align}
where we have defined the Nambu spinor $\hat{\Phineg}_{\eta k}=(\hat{f}^{\dag}_{\eta k\uparrow}~~\hat{f}_{\eta k\downarrow}
~~\hat{f}^{\dag}_{\eta k\downarrow}~~\hat{f}_{\eta k\uparrow})^{\dag}$ where $\hat{f}^{\dag}_{\eta k\sigma}$ and $\hat{f}_{\eta k\sigma}$ creates an electron and a hole, respectively with spin $\sigma$ and wave-vector $k$ in the ferromagnet $F_{\eta}$, $\eta=1,2$. The matrix $\hat{\mathbf{E}}_{\eta,k}(\theta)$ is written in the $4\times4$ Nambu space resulting from the tensor product between electron-hole and spin spaces. The general form of $\hat{\mathbf{E}}_{\eta,k}(\theta)$ is given by:
\begin{widetext}
\begin{align}\label{Etheta}
\hat{\mathbf{E}}_{\eta,k}(\theta)=
\begin{pmatrix}
   \epsilon_{k}-h_{\eta}\cos\theta-\mu_{_{\eta}}  &                     0               &      -h_{\eta}\sin\theta        & 0 \\
                0                    & -(\epsilon_{k}+h_{\eta}\cos\theta-\mu_{_{\eta}}) &                          & h_{\eta}\sin\theta \\
        -h_{\eta}\sin\theta                 &                     0               & \epsilon_{k}+h_{\eta}\cos\theta-\mu_{_{\eta}} & 0 \\
                0                    &              h_{\eta}\sin\theta            &              0           & -(\epsilon_{k}-h\cos\theta-\mu_{_{\eta}})
\end{pmatrix}.
\end{align}
\end{widetext}

The ferromagnets are modeled by the Stoner model\cite{fazekas} in which the spin bands of $F_{\eta}$ are split by an internal mean-field $h_{\eta}$ producing a finite polarization of the electron gas. The magnetization of $F_{1}$ is considered to point to a fixed direction while the magnetization of $F_{2}$ can be rotated by an arbitrary angle $\theta$.  The chemical potentials of each ferromagnet are determined by an external voltage bias $\mu_{\eta}=eV_{\eta}$ which controls the Fermi level of each electrode independently.

The superconductor is considered to be a conventional superconductor (\emph{s}-wave) being well described by the BCS Hamiltonian\cite{BCS}. In the Nambu notation this Hamiltonian reads:
\begin{align}\label{model:SC}
\hat{\mathcal{H}}_{S}=\sum_{k}\hat{\Phineg}^{\dag}_{sk}\hat{\mathbf{E}}_{S,k}\hat{\Phineg}_{sk}
\end{align}
with $\hat{\Phineg}_{sk}=(\hat{s}^{\dag}_{k\uparrow}~~\hat{s}_{k\downarrow}~~\hat{s}^{\dag}_{k\downarrow}~~\hat{s}_{k\uparrow})^{\dag}$
and
\begin{align*}
\hat{\mathbf{E}}_{S,k}=
\begin{pmatrix}
 \epsilon_{k}-\mu_{S}                & \Delta^{*} &         0               &        0               \\
  \Delta &         -(\epsilon_{k}-\mu_{S})        &         0               &        0               \\
  0                      &            0           &     (\epsilon_{k}-\mu_{S})               & -\Delta^{*} \\
  0                      &            0           & -\Delta  &       -(\epsilon_{k}-\mu_{S})
\end{pmatrix}.
\end{align*}

The superconducting correlations enter by means of the pair amplitude $\Delta$ which in general is a complex number depending on $k$. Since we are using just one superconductor lead, we use the well known assumption\cite{claro,claro2,principal} in which $\Delta$ is just a constant real number.  In addition, the superconductor chemical potential is fixed to zero as the ground ($\mu_{S}=0$).

The quantum dot is considered to be interacting with one level degenerated in spin,
\begin{align}\label{Hdqd}
\hat{\mathcal{H}}_{\mathcal{C}}=\mathbf{\hat{\Psi}}^{\dag}_{d}\mathbf{\hat{E}}_{d}\mathbf{\hat{\Psi}}_{d}+\mathcal{U}\hat{n}_{d\uparrow}\hat{n}_{d\downarrow}
\end{align}
where $\hat{\Psineg}_{d}=(\hat{d}^{\dag}_{\uparrow}~~\hat{d}_{\downarrow}~~\hat{d}^{\dag}_{\downarrow}~~\hat{d}_{\uparrow})^{\dag}$, and
\begin{align*}
\hat{\mathbf{E}}_{d}=
\begin{pmatrix}
  \varepsilon_{d} & 0 & 0 & 0 \\
  0 & -\varepsilon_{d} & 0 & 0 \\
  0 & 0 & \varepsilon_{d} & 0 \\
  0 & 0 & 0 & -\varepsilon_{d} \\
\end{pmatrix}.
\end{align*}

We consider that the QD level can be displaced by means of a gate voltage $V_{g}$, thus, $\varepsilon_{d}=\varepsilon_{0}-eV_{g}$ with $\varepsilon_{0}$ being the bare QD level (spin degenerated). The Coulomb correlations are described by $\mathcal{U}\hat{n}_{d\uparrow}\hat{n}_{d\downarrow}$ whose intensity is controlled by $\mathcal{U}$ which is considered to be smaller than the superconductor gap $\Delta$.

The tunneling between the QD and the leads is described by
\begin{align}\label{model:tunneling}
\hat{\mathcal{H}}_{T}
=
\sum_{k\gamma}[\hat{\Phineg}^{\dag}_{\gamma k}\hat{\mathbf{V}}_{\gamma k}\mathbf{\hat{\Psi}}_{d}+\mathbf{\hat{\Psi}}^{\dag}_{d}\hat{\mathbf{V}}^{\dag}_{\gamma k}\hat{\Phineg}_{\gamma k}]
\end{align}
in which
\begin{align*}
\hat{\mathbf{V}}_{\gamma k}=
\begin{pmatrix}
  V_{\gamma k} & 0 & 0 & 0 \\
  0 & -V^{*}_{\gamma k} & 0 & 0 \\
  0 & 0 & V_{\gamma k} & 0 \\
  0 & 0 & 0 & -V^{*}_{\gamma k} \\
\end{pmatrix}
\end{align*}
where $\gamma=1,2,s$ is the tunneling amplitude. Since the energy range is limited to the narrow superconductor gap, it is a good approximation to consider $\hat{\mathbf{V}}_{\gamma k}$  independent on $k$.

\subsection{Green's functions}

In order to calculate the transport properties we have used the
non-equilibrium Green's function method \cite{rammer}. All the physical
quantities can be cast in terms of the Green's function of the QD. In terms of Nambu spinors, the ``lesser'' ($\mathbf{G}^{<}$) and
retarded/advanced Green's function ($\mathbf{G}^{r/a}$) of the QD are written as
\begin{align}  \label{model:eq6a}
\mathbf{G}^{<}(t_{1},t_{2})=i\langle\hat{\boldsymbol{\Psi}}%
_{d}(t_{1})\otimes\hat{\boldsymbol{\Psi}}^{\dagger}_{d}(t_{2})\rangle
\end{align}
and
\begin{multline}  \label{model:eq6b}
\mathbf{G}^{r/a}(t_{1},t_{2})=\mp i\vartheta(\pm t_{1}\mp t_{2})\langle%
\hat{\boldsymbol{\Psi}}_{d}(t_{1})\otimes\hat{\boldsymbol{\Psi}}%
^{\dagger}_{d}(t_{2}) \\
+\hat{\boldsymbol{\Psi}}^{\dagger}_{d}(t_{2})\otimes\hat{\boldsymbol{\Psi}}%
_{d}(t_{1})\rangle,
\end{multline}
where the symbol $\otimes$ denotes a tensor product. Similar definitions are given for the leads Green's functions which can be expressed in terms of Eqs. \eqref{model:eq6a} and \eqref{model:eq6b}.

By using the equation of motion approach technique, along with the mean-field approximation (discussed in Appendix section), we obtain the Dyson's equation for the retarded Green's function:
\begin{align}\label{retarded}
\mathbf{G}^{r/a}(\varepsilon)
=
\mathbf{g}^{r/a}(\varepsilon)+\mathbf{g}^{r/a}(\varepsilon)\boldsymbol{\Sigma}^{r/a}(\varepsilon)\mathbf{G}^{r/a}(\varepsilon)
\end{align}
where $\mathbf{g}^{r/a}$ is the Green's function for the non-interacting QD isolated from the leads. It is written as
\begin{align*}
\mathbf{g}^{r/a}
=
\begin{pmatrix}
(x-\varepsilon_{d})^{-1} &   0   &   0   &   0
\\
0&   (x+\varepsilon_{d})^{-1}   &   0   &   0
\\
0   &   0   &   (x-\varepsilon_{d})^{-1} &0
\\
0 &   0   &   0   &   (x+\varepsilon_{d})^{-1}
\end{pmatrix}
\end{align*}
where we have defined $x=\varepsilon\pm i\eta$ and $\varepsilon_{d}=\varepsilon_{0}-eV_{g}$.

The self-energy $\mathbf{\Sigma}^{r/a}$ is given by
\begin{align}\label{selfenergy:ra}
\mathbf{\Sigma}^{r/a}(\varepsilon)=\mathbf{\Sigma}^{r/a}_{0}(\varepsilon)+\boldsymbol{\Theta}
\end{align}
where $\boldsymbol{\Theta}$ encodes the electronic correlations and $\mathbf{\Sigma}^{r}_{0}$ models the coupling between the QD and leads, i.e.,
\begin{align*}
\mathbf{\Sigma}^{r}_{0}(\varepsilon)=\mathbf{\Sigma}^{r}_{s}(\varepsilon)+\mathbf{\Sigma}^{r}_{1}(\varepsilon)+\mathbf{\Sigma}^{r}_{2}(\varepsilon)
\end{align*}
with
\begin{align}\label{SigmaRNr}
\boldsymbol{\Sigma}^{r}_{s}(\varepsilon)=-\dfrac{i}{2}\Gamma_{s}\varrho(\varepsilon)
\begin{pmatrix}
 1  & -\Delta/\varepsilon         & 0                 & 0 \\
 -\Delta/\varepsilon   & 1     & 0                 & 0 \\
  0             & 0                 & 1     & \Delta/\varepsilon \\
  0             & 0                 & \Delta/\varepsilon     & 1
\end{pmatrix}
\end{align}
modelling the coupling to the superconductor where $\Gamma_{s}=2\pi|V_{s}|^{2}\mathcal{D}_{s}(\varepsilon_{F})$ with $\mathcal{D}_{s}(\varepsilon_{F})$ being the density of states of the superconductor at the normal state solved at the Fermi level and $V_{s}$ is tunneling amplitude. We have also defined the generalized superconductor density of states,
\begin{align*}
\varrho(\varepsilon)=\dfrac{|\varepsilon|\vartheta(\varepsilon-\Delta)}{\sqrt{\varepsilon^{2}-\Delta^{2}}}-\dfrac{i\varepsilon\vartheta(\Delta-|\varepsilon|)}{\sqrt{\Delta^{2}-\varepsilon^{2}}}.
\end{align*}
in which the first term is the conventional BCS density of states\cite{BCS}, $\tilde{\varrho}(\varepsilon)=\text{Re}[\varrho(\varepsilon)]$ and the second term accounts for the Andreev bound states corresponding to evanescent waves representing the conversion of quasiparticles into Cooper pairs within the superconductor\cite{btk}.

Next, we define the self-energy due to the coupling with $F_{1}$:
\begin{align}\label{Sigma2}
\boldsymbol{\Sigma}_{1}^{r}(\varepsilon)=-\dfrac{i}{2}
\begin{pmatrix}
\Gamma_{1\uparrow} & 0 & 0 & 0
\\
0 & \Gamma_{1\downarrow} & 0 & 0
\\
0 & 0 &  \Gamma_{1\downarrow} & 0
\\
    0 & 0 & 0 & \Gamma_{1\uparrow}
\end{pmatrix}
\end{align}
with $\Gamma_{1\sigma}=2\pi |V_{1}|^{2}\mathcal{D}_{1\sigma}(\varepsilon_{F})$ where $V_{1}$ the hopping term and $\mathcal{D}_{1\sigma}(\varepsilon_{F})$ is the density of states per spin at the $F_{1}$ Fermi level.

The coupling with $F_{2}$ exhibits a similar form, however, the $F_{2}$ quantization axis is rotated by an angle $\theta$:
\begin{equation}  \label{sigmaF}
\mathbf{\Sigma}_{2}^{r,a}(\varepsilon)=\mp\frac{i}{2} \left(%
\begin{array}{cccc}
A_{\uparrow} & 0 & B & 0 \\
0 & A_{\downarrow} & 0 & B \\
B & 0 & A_{\downarrow} & 0 \\
0 & B & 0 & A_{\uparrow}%
\end{array}%
\right),
\end{equation}
with $A_{\sigma}\equiv c^{2}\Gamma_{2\sigma}+s^{2}\Gamma_{2%
\bar{\sigma}}$, $B=sc(\Gamma_{2\uparrow}-\Gamma_{2\downarrow})$, $%
s\equiv\sin\theta/2$ and $c\equiv\cos\theta/2$. We also have defined $\Gamma_{2\sigma}=2\pi |V_{2}|^{2}\mathcal{D}_{2\sigma}(\varepsilon_{F})$ where $V_{2}$ the hopping term and $\mathcal{D}_{2\sigma}(\varepsilon_{F})$ is the density of states per spin at the $F_{2}$ Fermi level. It is worth mentioning that we have taken the wide-band limit in which the $\Gamma_{1}$, $\Gamma_{2}$ and $\Gamma_{s}$ are assumed to be constants.\\


\subsection{Physical Quantities}
In this section we derive the physical quantities and the relevant parameters used in the analysis of the results presented in Sec. \ref{resdic}.

\subsubsection{Ferromagnet Polarization}

Within the Stoner model, the electrons gas polarization is a result of the exchange mean field due to the electron-electron interaction. In this, way we define the polarization for the ferromagnet $F_{\alpha}$ as follows:
\begin{align}\label{pol}
P_{\alpha}=\dfrac{\Gamma_{\alpha\uparrow}-\Gamma_{\alpha\uparrow}}{\Gamma_{\alpha\uparrow}+\Gamma_{\alpha\uparrow}}
\end{align}
where $\alpha=1,2$.  Here, the coupling constants $\Gamma_{\alpha\sigma}$ are considered as independent parameters.

%

\subsubsection{Electrical Current}
By using the equation of motion method, one is able to calculate the current between the ferromagnet $F_{\alpha}$ and the QD. By using the time variation of the average occupation of the lead $F_{\alpha}$ we obtain the following equation:
\begin{align}\label{apendiceD:correnteI1:def}
I_{\alpha}=\frac{e}{h}\int d\varepsilon\left[
\mathbf{G}^{r}(\varepsilon)\mathbf{\Sigma}_{\alpha}^{<}(\varepsilon)
+\mathbf{G}^{<}(\varepsilon)\mathbf{\Sigma}_{\alpha}^{a}(\varepsilon)+\text{H.c.}\right]_{11+33}
\end{align}
where  $\alpha=1,2$. The current $I_{1}$ is explicitly written as
\begin{align}\label{I1}
I_{1}=I_{12}+I_{1s}
\end{align}
where we have defined the co-tunneling current as
\begin{align}\label{I12}
I_{12}=\frac{e}{h}\int T_{12}(f_{1}-f_{2}),
d\varepsilon,
\end{align}
and the current flowing between $F_{1}$ and the superconductor is given by
\begin{align}\label{I1s}
I_{1s}=\frac{e}{h}\int
\left[T_{AR, 11}(f_{1}-\bar{f_{1}})+T_{AR,12}(f_{1}-\bar{f_{2}})\right]~d\varepsilon,
\end{align}
where $f_{1}$ and $f_{2}$ are the corresponding Fermi distributions for electrons in the leads $F_{1}$ and $F_{2}$ and $\bar{f}_{1}$ and $\bar{f}_{2}$ are the corresponding ones for holes. By comparing the Fermi distributions one is able to determine each contribution in Eq. \eqref{I1}. In fact, $T_{12}$ is the co-tunneling current of electrons from $F_{1}$ to $F_{2}$ through the QD; $T_{AR, 11}$ accounts for the Andreev reflection in $F_{1}$ and finally $T_{AR, 12}$ is the crossed Andreev reflection of an electron from $F_{1}$ as a hole in $F_{2}$. The transmittance expressions for each process are given by
\begin{widetext}
\begin{subequations}\label{amplitudes2}
\begin{align}
\label{a11m}
T_{AR, 11}=\Gamma_{1\uparrow}\left(|G^{r}_{14}|^{2}\Gamma_{1\uparrow}+|G^{r}_{12}|^{2}\Gamma_{1\downarrow}\right)
+
\Gamma_{1\downarrow}\left(|G^{r}_{34}|^{2}\Gamma_{1\uparrow}+|G^{r}_{32}|^{2}\Gamma_{1\downarrow}\right)
\end{align}
\begin{align}\label{a12m}
T_{AR, 12}=\Gamma_{1\uparrow}[
(c^{2}\Gamma_{2\uparrow}+s^{2}\Gamma_{2\downarrow})|G^{r}_{14}|^{2}
+
(s^{2}\Gamma_{2\uparrow}+c^{2}\Gamma_{2\downarrow})|G^{r}_{12}|^{2}
+sc(\Gamma_{2\uparrow}-\Gamma_{2\downarrow})([G^{r}_{12}]^{*}G^{r}_{14}
+[G^{r}_{14}]^{*}G^{r}_{12})]
\\\nonumber
+\Gamma_{1\downarrow}[(c^{2}\Gamma_{2\uparrow}+s^{2}\Gamma_{2\downarrow})|G^{r}_{34}|^{2}+(s^{2}\Gamma_{2\uparrow}+
c^{2}\Gamma_{2\downarrow})|G^{r}_{32}|^{2}
+sc(\Gamma_{2\uparrow}-\Gamma_{2\downarrow})([G^{r}_{32}
]^{*}G^{r}_{34}+[G^{r}_{34}
]^{*}G^{r}_{32})]
\end{align}
\begin{align}\label{q12m}
T_{12}=\Gamma_{1\downarrow}[(s^{2}\Gamma_{2\uparrow}+c^{2}\Gamma_{2\downarrow})|G^{r}_{33}|^{2}
+(c^{2}\Gamma_{2\uparrow}+s^{2}\Gamma_{2\downarrow})|G^{r}_{31}|^{2}
+sc(\Gamma_{2\uparrow}-\Gamma_{2\downarrow})([G^{r}_{33}
]^{*}G^{r}_{31}+[G^{r}_{31} ]^{*}G^{r}_{33})]
\\\nonumber+\Gamma_{1\uparrow}[(s^{2}\Gamma_{2\uparrow}+c^{2}\Gamma_{2\downarrow})|G^{r}_{13}|^{2}+
(c^{2}\Gamma_{2\uparrow}+s^{2}\Gamma_{2\downarrow})|G^{r}_{11}|^{2}
+sc(\Gamma_{2\uparrow}-\Gamma_{2\downarrow})([G^{r}_{13}
]^{*}G^{r}_{11}+[G^{r}_{11} ]^{*}G^{r}_{13})].
\end{align}
\end{subequations}
\end{widetext}
The corresponding equation for $I_{2}$ can be obtained from $I_{1}$ just replacing the $1\rightarrow2$ in the previous equations. We point out that the expression for the current is the same as obtained by Y. Zhu \textit{et. al.}\cite{principal} for a noninteracting QD. In the present case, in spite from the fact of the current formula resembles the one obtained in Ref. \onlinecite{principal}, it is being considered the presence of interactions into the QD which means that the matrix elements of the Green's function must be determined in a self-consistent calculation, due to Eq. \eqref{Theta}. However, within the approximation used in this work,  it is still possible to obtain a Landauer-like equation for the current. This is an advantage in the sense that one can obtain analytic expressions for the transmittances $T_{12}$, $T_{AR,11}$ and $T_{AR,12}$.

\section{\label{resdic}Results and Discussion}

In the following results, we consider the Andreev regime in which the applied bias range is bounded by the superconductor energy gap. Since this quantity is the natural energy scale of the problem, all the physical parameters are presented in units of the superconductor gap. We start with the zero-bias regime and then the finite-bias case is considered.

\subsection{Zero-bias regime}

In order to clarify the effects of the electronic correlation within the QD on transport properties, we consider the zero-bias regime firstly and analyze the role of the interactions appearing in the self-energy, Eq. \eqref{selfenergy:ra}. As show in the following results, the mean-field approximation just renormalizes the QD energy level developing the same role in the system as the gate voltages.

\subsubsection{Zero-Bias Transmittance for $F_{1}-QD-F_{2}$ system}

\begin{figure}[!]\centering
\includegraphics[scale=.85]{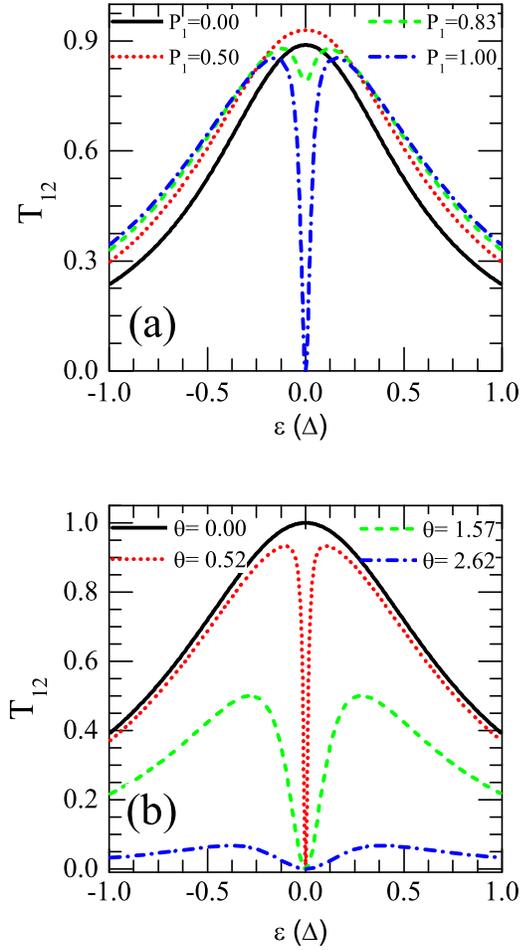}
\caption{(Color Online)~\label{fig2}Transmittance curves ($T_{12}$) for the system $F_{1}-QD-F_{2}$, i.e., in the absence of the superconductor lead. (a) $T_{12}$ curves for different polarization values $P_{1}$ of the ferromagnet $F_{1}$ while the magnetization of $F_{2}$ is aligned at an angle $\theta=\pi/4$ with the respect to the magnetization of $F_{1}$. (b) $T_{12}$ curves with $P_{1}$ fixed to unity and changing the magnetization angle of $F_{2}$ from a parallel alignment $\theta=0$ towards a orientation close to $\theta=\pi$ where $T_{12}=0$, see Eq. \eqref{T12:Scdecoupled}. Fixed parameters:  $V_{1}=V_{2}=0$, $\Gamma_{1}=0.40$, $\Gamma_{2}=0.40$, $\Gamma_{s}=0$ and $P_{2}=1.0$. All the parameters are scaled by the energy gap of the superconductor lead.}
\end{figure}

We consider the transmittance for electrons between the ferromagnetic leads through the QD. In this case, we start with simplest case in which the electronic correlations are absent. In Fig. \ref{fig2} it is shown the effect of the magnetization on the electronic transport. By setting the relative angle between the magnetizations of $F_{1}$ and $F_{2}$ to $\theta=\pi/4$, an intermediate angle, we have calculated the transmittance for different values of $P_{1}$, the polarization of $F_{1}$. In Fig. \ref{fig2}a, for $P_{1}=0$ the transmittance curve is just a resonance whose width is determined by the hybridization between the discrete level of the QD with the continuum of states from the ferromagnet bands. When $P_{1}$ is increased a sharp dip emerges for $\varepsilon=0$ whose height increases with $P_{1}$. For $P_{1}=1$ this dip reaches the horizontal axis and the transmittance is zero for $\varepsilon=0$.
\begin{figure*}[!]\centering
\includegraphics[scale=2.5]{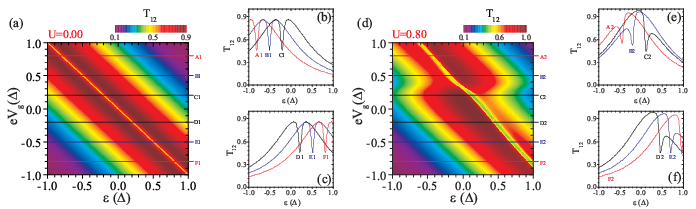}
\caption{\label{fig3}(Color Online)~Transmittance curves for the system $F_{1}-QD-F_{2}$, i.e., in the absence of the superconductor lead. (a)~Contour plot for zero-bias transmittance $T_{12}$ in terms of the gate potential $V_{g}$ and energy $\varepsilon$ for $\mathcal{U}=0$. (b) $T_{12}$ curves for positive values of $V_{g}$. Their location at the contour plot are indicated by the horizontal lines labeled by A1, B1 and C1 for $V_{g}$ equal to  0.8, 0.5 and 0.2, respectively. (c)~$T_{12}$~curves for negative values of $V_{g}$ whose location in the contour plot is given by D1, E1 and F1 lines for $V_{g}$ equal to -0.2, -0.5 and -0.8, respectively.~(d) Contour plot for zero-bias transmittance $T_{12}$ in terms of the gate potential $V_{g}$ and energy $\varepsilon$ for $\mathcal{U}=0.8$. (e)~$T_{12}$ curves with A2, B2 and C2 corresponding to $V_{g}$ equal to 0.8, 0.5 and 0.2, respectively. (f) $T_{12}$ curves for negative gate voltage values with D2, E2 and F2 corresponding to $V_{g}$ equal to -0.2, -0.5 and -0.8, respectively. Fixed parameters: $\theta=\pi/4$, $V_{1}=V_{2}=0$, $\Gamma_{1}=0.40$, $\Gamma_{2}=0.40$, $\Gamma_{s}=0$, $P_{1}=0.95$ and $P_{2}=1.0$. All the parameters are scaled by the energy gap of the superconductor lead.}
\end{figure*}

In Fig. \ref{fig2}b the evolution of the dip is studied by varying the angle $\theta$ while the polarization $P_{1}$ is fixed to the unity as the limit case in which the dip exhibits the most pronounced size. For $\theta=0$ the resonance behavior is recovered but as long as $\theta$ is different of zero the dip appears and the transmittance is pushed to zero at $\varepsilon=0$.  As $\theta$ increases towards $\pi$ the transmittance is suppressed and the dip opens up revealing a two peak structure which is illustrated by the curves for $\theta=1.57$ and $\theta=2.62$. Notice that the transmittance is zero in the whole range when $\theta=\pi$ and the ferromagnets are full polarized. The zero-bias results of Fig. \ref{fig2} are easily explained by considering the expression for the transmittance $T_{12}$ given by Eq. \eqref{q12m} where it can be noted that the effect of the ferromagnetism is to create two interfering channels for spins up and down. This interference pattern resulting in the transmittance curves of Fig. \ref{fig2} is dependent on the polarizations $P_{1}$, $P_{2}$ and the angle $\theta$. Additionally, the matrix elements of the retarded Green's function encode the processes in which the electron is scattered within the QD. To illustrate this point, we notice that the off-diagonal elements $G^{r}_{13}$ and $G^{r}_{31}$ represent a spin-flip process into the QD due to the misalignment of the magnetization of the ferromagnets. As $\theta$ is set to $0$ or $\pi$, while the ferromagnets are full polarized, these contributions are removed and the interference effect is completely suppressed. In this case, the transmittance is either a maximum (when $\theta=0$) or 0  (when $\theta=\pi$) when there are no states available for electrons in both magnets.

Next, we consider the effect of the gate voltage on the transmittance curves.  In Fig. \ref{fig3}, a contour plot for transmittance in terms of the energy $\varepsilon$ and the gate voltage, $V_{g}$, is shown. In order to explore the interference effect related to the different spin channels of conduction, we have used $P_{1}=0.95$ and $\theta=\pi/4$ which leads to a small dip on the transmittance. The resulting contour plot exhibits a well localized diagonal line connecting the points $(\varepsilon=1, eV_{g}=-1)$ and $(\varepsilon=-1, eV_{g}=+1)$. This means that the effect of the gate voltage is just displace the point at which the completely destructive interference occurs.  In fact, this behavior can again be understood by considering the expression for $T_{12}$ and noting that the gate voltage just renormalizes the QD level. In particular, for the full polarized case, it is possible to derive a rather compact transmittance expression by substituting the corresponding Green's functions matrix elements into Eq. \eqref{q12m}. After some algebra, one ends up with the following expression:
\begin{figure*}[!]\centering
\includegraphics[scale=2.5]{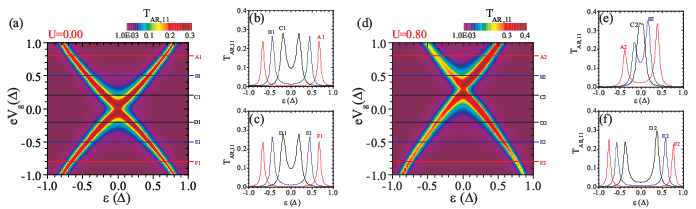}
\caption{\label{fig4}(Color Online)~Transmittance curves for the system $F_{1}-(QD,S)-F_{2}$ corresponding to direct Andreev reflection at ferromagnet $F_{1}$, $T_{AR,11}$. (a)~Contour plot for zero-bias Andreev transmittance $T_{AR,11}$ in terms of the gate potential $V_{g}$ and energy $\varepsilon$ for $\mathcal{U}=0$. (b) $T_{AR,11}$ curves for positive values of $V_{g}$. Their location at the contour plot are indicated by the horizontal lines labeled by A1, B1 and C1 for $V_{g}$ equal to  0.8, 0.50 and 0.2, respectively. (c)~$T_{AR,11}$~curves for negative values of $V_{g}$ whose location in the contour plot is given by D1, E1 and F1 lines for $V_{g}$ equal to -0.2, -0.5 and -0.8, respectively.~(d) Contour plot for zero-bias transmittance $T_{AR,11}$ in terms of the gate potential $V_{g}$ and energy $\varepsilon$ for $\mathcal{U}=0.8$. (e)~$T_{AR,11}$ curves with A2, B2 and C2 corresponding to $V_{g}$ equal to 0.8, 0.5 and 0.2, respectively. (f) $T_{AR,11}$ curves for negative gate voltage values with D2, E2 and F2 corresponding to $V_{g}$ equal to -0.2, -0.5 and -0.8, respectively. Fixed parameters: $\theta=\pi/4$, $V_{1}=V_{2}=0$, $\Gamma_{1}=0.40$, $\Gamma_{2}=0.40$, $\Gamma_{s}=0.40$, $P_{1}=0.95$ and $P_{2}=1.0$. All the parameters are scaled by the energy gap of the superconductor lead.}
\includegraphics[scale=2.5]{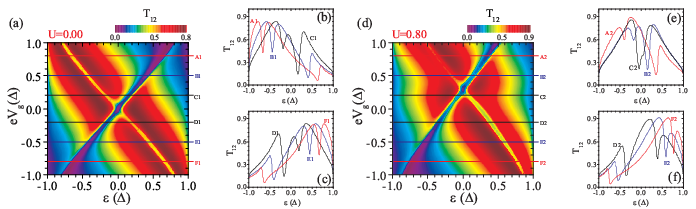}
\caption{\label{fig5}(Color Online)~Transmittance curves for the system $F_{1}-(QD,S)-F_{2}$, i.e., with the presence of the superconductor lead coupled to the QD. (a)~Contour plot for zero-bias transmittance $T_{12}$ in terms of the gate potential $V_{g}$ and energy $\varepsilon$ for $\mathcal{U}=0$. (b) $T_{12}$ curves for positive values of $V_{g}$. Their location at the contour plot are indicated by the horizontal lines labeled by A1, B1 and C1 for $V_{g}$ equal to  0.8, 0.50 and 0.2, respectively. (c)~$T_{12}$~curves for negative values of $V_{g}$ whose location in the contour plot is given by D1, E1 and F1 lines for $V_{g}$ equal to -0.2, -0.5 and -0.8, respectively.~(d) Contour plot for zero-bias transmittance $T_{12}$ in terms of the gate potential $V_{g}$ and energy $\varepsilon$ for $\mathcal{U}=0.8$. (e)~$T_{12}$ curves with A2, B2 and C2 corresponding to $V_{g}$ equal to 0.8, 0.5 and 0.2, respectively. (f) $T_{12}$ curves for negative gate voltage values with D2, E2 and F2 corresponding to $V_{g}$ equal to -0.2, -0.5 and -0.8, respectively. Fixed parameters: $\theta=\pi/4$, $V_{1}=V_{2}=0$, $\Gamma_{1}=0.40$, $\Gamma_{2}=0.40$, $\Gamma_{s}=0.40$, $P_{1}=0.95$ and $P_{2}=1.0$. All the parameters are scaled by the energy gap of the superconductor lead.}
\end{figure*}
\begin{align}\label{T12:Scdecoupled}
T_{12}(\tilde{\varepsilon})=
\dfrac{4\tilde{\varepsilon}^{2}\widetilde{\Gamma}^{2}(\theta)}
{[\widetilde{\Gamma}(\theta-\pi/2)]^{4} +\tilde{\varepsilon}^{2}[\Gamma_{1}^{2}+2\widetilde{\Gamma}^{2}(\theta)+\Gamma_{2}^{2}+\tilde{\varepsilon}^{2}]}
\end{align}
where $\widetilde{\Gamma}(\theta)=\sqrt{\Gamma_{1}\Gamma_{2}} \cos(\theta/2)$, $\tilde{\varepsilon}=\varepsilon-\varepsilon_{d}$,  $\varepsilon_{d}=\varepsilon_{0}-eV_{g}$ with $\varepsilon_{0}$ being the bare QD level and $\Gamma_{i}=(\Gamma_{1\uparrow}+\Gamma_{i\downarrow})/2$, $i=1,2$ are the spin averaged couplings of the QD with $F_{1}$ and $F_{2}$.  By setting the condition $T_{12}=0$ one obtains that the dip is located at $\varepsilon=-eV_{g}$ where we have considered the bare level of the QD, $\varepsilon_{0}=0$ and the electronic charge constant $e>0$. Accordingly, the equation $\varepsilon=-eV_{g}$ describes the diagonal line that locates the dip in the transmittance in Fig. \ref{fig3}a. For polarization values slight smaller than unity, there would be constant added to right-hand side of  $\varepsilon=-eV_{g}$ which leads to a dip with height smaller than as the one shown in the contour plot. However, the minimum value of $T_{12}$ is still located as in the full polarized case. In Figs. \ref{fig3}b and \ref{fig3}c some representative curves of $T_{12}$ are illustrated whose location in the contour plot is given by the horizontal lines labeled by \textsf{A1}, \textsf{B1} and \textsf{C1} for negative values of $eV_{g}$ and \textsf{E1}, \textsf{F1} and \textsf{G1} for the corresponding negative values of $eV_{g}$. It can be noted that the gate just displaces the dip along the $\varepsilon$-axis and the curve still carries a symmetric profile with respect to the dip. In this way, the effect of the gate voltage is just to produce a rigid displacement of the point at which the destructive interference between the spin channels occurs.

In Fig. \ref{fig3}d it is shown the transmittance $T_{12}$ contour plot under the presence of electronic correlations at the QD. The strength of these correlations is given by the parameter $\mathcal{U}=0.80$ in superconductor gap units. In this case, $T_{12}$ is dependent on the occupation of the QD for both spins and on the spin-flip averages of the form $\bra\hat{d}_{\sigma}^{\dag}\hat{d}_{\bar{\sigma}}\ket$ with $\sigma=\uparrow, \downarrow$ and $\bar{\sigma}=-\sigma$, c.f. Eq. \eqref{Theta}. As a result, the symmetry with respect to the sign of both, the gate voltage $V_{g}$ and energy $\varepsilon$ is broken as evident from the contour plot in Fig. \ref{fig3}d. The correlations enter into the expressions for Green's functions through Eq. \eqref{Theta} which leads to a self-consistent calculation by means of the Keldysh equation given by Eq. \eqref{lesser:final:def2} (see Appendix).   In Figs. \ref{fig3}e and \ref{fig3}f some representative curves are also illustrating an additional lack of symmetry on the transmittance curves. The adjacent peaks located at each side of the dip are now presenting different heights. As evident from the curves \textsf{A2} and \textsf{B2} the right peak is higher and wider in comparison to the left peak. This trend is inverted for $V_{g}\sim0.32$ where the right peak is suppressed and transmittance exhibits a higher value for the left peak. This behavior is maintained from larger values of $V_{g}$ as one can see from the curves \textsf{C2} up to \textsf{G2}. These results show that the interaction on the QD just provides minor changes on the transport properties within the mean-field approximation used in this work. In fact the physics is ruled by the coupling constants appearing in Eq. \eqref{q12m} which moderate the role of each matrix element of the Green's function of the QD.

\subsubsection{Zero-Bias Transmittance for $F_{1}-(QD,S)-F_{2}$ system}

Next, we consider the full system with the presence of the superconductor lead coupled to the QD as illustrated in Fig. \ref{fig1}. It is worth recalling  that we are interested in the Andreev conduction regime in which all the parameters are restrict to energies within the superconductor gap. In this range of energy, the superconductor acts as a barrier which rules out the direct tunneling of quasi-particles from the leads to the superconductor. In spite of this restrictive condition, it is still possible to a current to take place from the leads to the superconductor by means of AR. In the electrical current equation, the direct and crossed ARs contributions are included by means of the transmittance expressions given by Eqs. \eqref{a11m} and \eqref{a12m} respectively. Notice that the Green's function matrix elements appearing in these expressions are related to conversion of an electron of spin $\sigma$ into a hole of spin $\bar{\sigma}$. Due to the coupling with the second ferromagnet, whose magnetization may be pointing in an arbitrary orientation,  processes involving spin-flip also contribute to the full transmittance.

To contrast the AR with the previous results for the $F_{1}-QD-F_{2}$ system, in Fig. \ref{fig4} is shown the direct Andreev transmittance $T_{AR,11}$ corresponding the process where an electron from $F_{1}$ is reflected as a hole in the same lead $F_{1}$. The pattern observed was obtained for $P_{1}=0.95$ and in Fig. \ref{fig4}a it is shown the dependence of $T_{AR,11}$ with the energy $\varepsilon$ and gate voltage $V_{g}$. As the gate voltage changes from zero, a  double peak structure emerges with the separation of the peaks increasing with $eV_{g}$. These two peaks represent the so-called Andreev bound states which are virtual states of superconducting quasi-particles formed by a pair of an electron with a hole which is converted into a Cooper pair as it enters inside the superconductor. These are strongly suppressed by the ferromagnetic polarization once the conventional superconductivity requires anti-parallel alignment of the electronic spins. As a result, in the limit of high polarization, the direct Andreev contribution has a minor contribution to the transport and is completely eliminated when $P_{j}=1$, $j=1,2$. This can be observed in Figs. \ref{fig4}b and \ref{fig4}c where the transmittance amplitude is confined to values around 0.3 for $P_{1}=0.95$.
\begin{figure*}[!]\centering
\includegraphics[scale=.53]{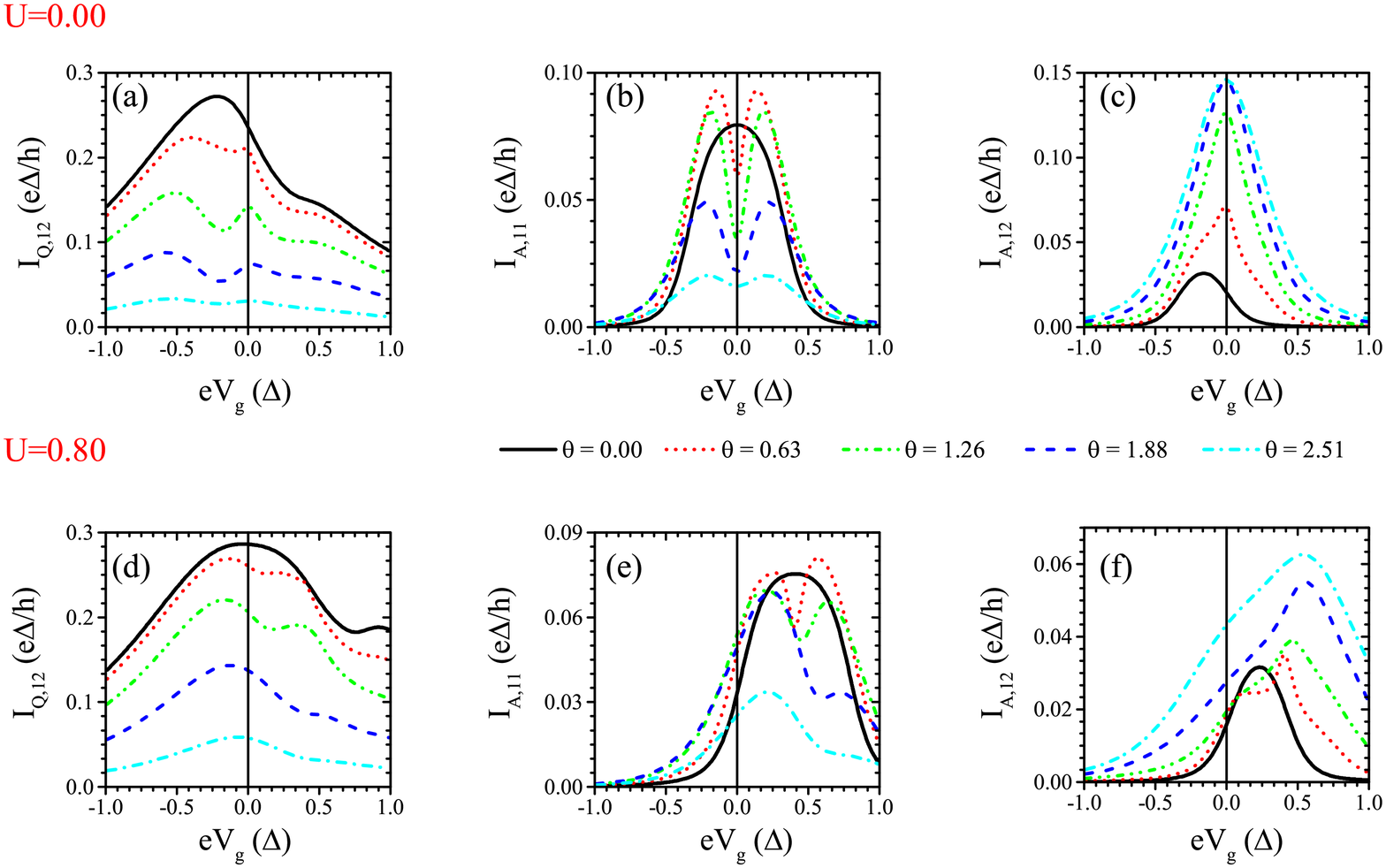}
\caption{(Color Online)~\label{fig6} Currents flowing through the lead $F_{1}$ for the system $F_{1}-(QD-S)-F_{2}$: $I_{Q,12}$ is the co-tunneling current, $I_{A,11}$ is the direct Andreev current and $I_{A,12}$ is the crossed Andreev current. It is considered a finite bias $V_{1}=0.30$ applied to $F_{1}$ while $F_{2}$ is kept grounded. The upper figures show the current profiles under the absence of Coulomb correlations within the QD. For intermediate values of $\theta$, the direct Andreev current, $I_{A,11}$ exhibits a dip at $V_{g}=0$ while a corresponding peak emerges in $I_{Q,12}$. A similar pattern is observed in Figs. (d) and (e) at $V_{g}\sim0.40$ for the interacting case $U=0.80$. Notice that for $\theta$ close to $\pi$ the conduction is ruled by the crossed AR process as shown in Figs. (c) and (f). Fixed parameters:  $V_{1}=0.30$,  $V_{2}=0$, $\Gamma_{1}=\Gamma_{2}=\Gamma_{s}=0.4$, $P_{1}=0.95$, $P_{2}=1.0$.   All the parameters are scaled by the energy gap of the superconductor lead.}
\end{figure*}

In Fig. \ref{fig4}d, the Andreev transmittance is shown for the interacting case in which the interaction strength $\mathcal{U}=0.80$. In this case, it can be noted that the transmittance  has a similar pattern as the noninteracting case except that the intersection point where the two peaks merge into a single one is shifted along the gate voltage axis. In addition, the height of the peaks are also different as one can observe by comparing the Figs. \ref{fig4}c and \ref{fig4}d with the curves shown in Figs. \ref{fig4}e and \ref{fig4}f. This asymmetry is stronger for positive values of the gate voltage as evident from the \textsf{A2} curve of Fig. \ref{fig4}e.  As shown in previous works \cite{siqueira,siqueira2}, this asymmetry is crucial for the transport since the Cooper pairs are formed by injecting two electrons with opposite spins and energy signs at once into the superconductor. In this way, the current being injected into the superconductor is determined by smaller peak of the transmittance.

Next, we consider the tunneling between the ferromagnets characterized by the transmittance $T_{12}$. The corresponding curves for both noninteracting and interacting cases are shown in Fig. \ref{fig5}. The main difference when the superconductor is coupled into the QD is the emergence of a second diagonal dip line in the contour plot shown in Fig. \ref{fig5}a. Thus, the contour plot is divided in four triangular regions in which the transmittance can reach a maximum for particular values of gate voltage and energy. The curves shown in Figs. \ref{fig5}b and \ref{fig5}c reveal a two dip structure in transmittance with a central well defined peak. In this way, the effect of the superconductor into $T_{12}$ is just to introduce a second state at which the channels of spins interfere  destructively. This is a signature of the Andreev bound states into the transport between the ferromagnets. In fact, by comparing the contour plots of Fig. \ref{fig5}a and  \ref{fig5}d with the corresponding ones of Fig. \ref{fig4}a and  \ref{fig4}d, it is clear that these are complementary patterns of resonances: in the regions at which the $T_{AR,11}$ exhibits a maximum value, the transmittance $T_{12}$ presents a dip. Thus, the coupling with the superconductor  results in a leakage of states from the direct channel between the ferromagnets for the Andreev states. This leads to the pattern observed in Fig. \ref{fig5}.  It is worth mentioning that Calle \text{et. al.}\cite{calle} have studied a three-terminal nanostructure composed by two normal metals coupled by a double quantum dot system and a superconductor. In this system they have observed a similar pattern as shown in Fig. \ref{fig5} with two dips and a central peak in the transmittance $T_{12}$. The authors attributed such a feature to the Fano effect induced by the second quantum dot. Here, the origin of such a pattern is related to the interplay between the Andreev bound states and spin polarization provided by the ferromagnets. As a result of these correlations, the spectral properties of the quantum dot are similar to the double quantum dot structure of Ref. \onlinecite{calle}.

\subsection{Finite-bias regime}

In the finite bias regime the correlations appearing in Eq. \eqref{Theta} couple the transmittance and local density of states (LDOS) with the bias applied to $F_{1}$ and $F_{2}$. In this way, for each value of $V_{1}$  and $V_{2}$ there is a corresponding transmittance and LDOS curve. In this way, the dependence of these quantities on the applied bias becomes more intricate than the zero-bias case.  In spite of these modifications, it is also possible to recognize the signatures of Fano interference in the non-equilibrium case. In order to illustrate such an effect, we have calculated the electrical current for a finite bias ($eV_{1}=0.30$) keeping the other parameters with the same values used in Fig. \ref{fig5}. In Fig. \ref{fig6}a  the co-tunneling current $I_{Q,12}$ is shown for some values of $\theta$,  with  $\mathcal{U}=0$. For $\theta=0$ (solid-black curve), the current reaches a maximum value for $eV_{g}=-0.25$ and then decreases for positive values of $eV_{g}$. As $\theta$ is increased the pattern changes with a second peak appearing for $eV_{g}=0$. This peak is well pronounced for $\theta=1.26$ (dot-dashed green curve) and starts being suppressed as $\theta>1.88$. In Figs. \ref{fig6}b and \ref{fig6}c the currents due to the direct and crossed AR are shown, respectively. In the direct AR, there is a corresponding dip at $eV_{g}=0$ corresponding to the peak appearing in the co-tunneling current. On the other hand, the crossed AR also presents a peak at $eV_{g}=0$ which increases with $\theta$. This behavior is a result of the high polarization values which suppressed the available states for local tunneling processes like the direct AR and co-tunneling from $F_{1}$ to $F_{2}$. As a result, the crossed AR is the dominant process for $\theta$ close to $\pi$ once electrons of opposite spins from different leads combine into a Cooper pair in $S$. This is evident by comparing the amplitude of $I_{Q,12}$, $I_{A,11}$ and $I_{A,12}$ for $\theta=2.51$. The peak appearing at $eV_{g}=0$ for $I_{Q,12}$ as $\theta$ is changed from 0 to $\pi$ is the non-equilibrium signature of the Fano-like interference appearing in the zero-bias curves. This effect is also present under the presence of Coulomb correlations within the QD. The corresponding curves for $\mathcal{U}=0.80$ are plotted in Figs. \ref{fig6}c, \ref{fig6}d and \ref{fig6}e. It can be noted that the presence of the peak in the co-tunneling and crossed AR currents are shifted to $eV_{g}\sim0.46$ for $\theta=2.51$. A corresponding dip in the direct AR is also present at the same point which illustrates the fact of the interaction, within the mean-field approximation, just shifts the resonance condition in the same form as the zero-bias case.


\subsubsection{Spin-degeneracy}
\begin{figure}[h]\centering
\includegraphics[scale=.84]{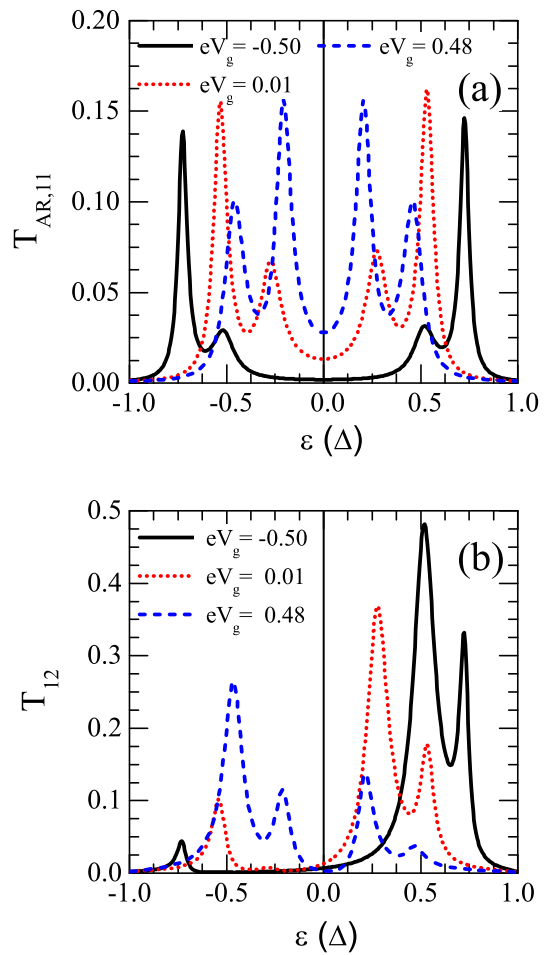}
\caption{(Color Online)~\label{fig6nw} Transmittance curves for finite bias. (a) Transmittance curves for direct Andreev reflection $T_{AR,11}$ (b) Co-tunneling transmittance $T_{12}$. By reducing the constant coupling to the ferromagnetic leads, it is possible to observe the splitting of the Andreev resonances due to the raising of spin degeneracy.  Fixed parameters:  $V_{1}=0.95$,  $V_{2}=0$, $\Gamma_{1}=0.10, \Gamma_{2}=0.05, \Gamma_{s}=1.0$, $P_{1}=0.85$, $P_{2}=1.0$, $\mathcal{U}=0.85$, $\theta=3\pi/2$, $k_{B}T=0.05$.   All the parameters are scaled by the energy gap of the superconductor lead.}
\end{figure}

In the zero-bias curves shown in Fig. \ref{fig4}, the transmittance curves exhibit a double peak structure related to the Andreev resonances. However, it is expected a splitting of these resonances due to the raising of the spin degeneracy caused by the Coulomb correlation within the QD. In Fig. \ref{fig6nw}a, it is shown the transmittance curves for both $T_{AR,11}$~and~$T_{12}$ for a finite bias voltage $V_{1}=0.95$. We also have chosen small values for the coupling to the ferromagnets, $\Gamma_{1}=0.1$ and $\Gamma_{2}=0.05$ which are crucial to allow the resolution of the spin degeneracy.  In this regime, it is possible to observe such a splitting of the peaks in which the $T_{AR,11}$ curves exhibit a four peak structure.

By changing the gate voltage, it is possible to change the pattern as one can observe by comparing the curves for $eV_{g}=-0.50$, $eV_{g}=0.01$ and $eV_{g}=0.48$. For negative values of the gate voltage, the central peaks are suppressed while for positive values the pattern is better resolved. The asymmetry with respect to the signal of the gate voltage is also a result of the interaction within the QD. This is clearer in the zero-bias regime in which the resonance condition is shifted by the presence of the interaction. A similar behavior is observed for the co-tunneling transmittance $T_{12}$, as illustrated in Fig. \ref{fig6nw}b. Notice that the $T_{12}$ curves do not present the corresponding Fano-like resonance as observed for zero-bias regime. In fact, the position of the peaks of $T_{12}$ in Fig. \ref{fig6nw}b are coincident with those of $T_{AR,11}$ in Fig. \ref{fig6nw}a. The values of the parameters to obtain the Fano-like resonance in the co-tunneling transmittances are different from those that allows for the resolution of the peaks due to the spin degeneracy. In this way, it is not possible to observe both effects with the same set of parameters.

\section{\label{concl}Conclusion}

In this work, we have studied the interference effects on $F_{1}-(QD,S)-F_{2}$ due to the coupling to a conventional superconductor. By varying the angle between the two magnetization it is possible to obtain a very pronounced dip in the transmittance for $\varepsilon=0$ when the superconductor is decoupled from the QD. In contrast, the interplay between spin imbalance and Andreev bound states gives rise to a central peak at $\varepsilon=0$ when the superconductor is coupled to the system. Such an effect is a result of the interference between the different channels of conduction through the QD.  Additionally, the leakage of states for Andreev transport also introduces two anti-resonances in the zero-bias transmittance for the co-tunneling of electrons between the ferromagnets.

The effects of correlations were taken within a generalized mean-field approximation also taking into account spin-flip correlations and proximity effect due to the coupling to the superconductor. Such correlations are relevant since the physical quantities must be determined in self-consistent way for each value of gate and bias voltages (for non-equilibrium situation) thus introducing a nontrivial dependence on these quantities. In fact, as shown in Figs. \ref{fig3} and \ref{fig5}, the combination of these correlations breaks the symmetry in the transmittance and shifts the region at which the Fano-like interference takes place. In the non-equilibrium situation, it is also possible to observe the signatures of such an interference in the electrical current as shown in Fig. \ref{fig6}.
The approximation scheme used in this work allows us to write the electrical current in a Landauer-like equation. In this way, it is possible to obtain analytic expressions for the transmittance for both Andreev and co-tunneling contributions. Additionally, we have restricted the calculations for large values of ferromagnetic polarization reducing in this way the fluctuation in the occupation numbers.  Under this condition, the approximation yields results in a good agreement with other approximations schemes.  The results above can be reproduced in experiments by using half-metal ferromagnets. Additionally, high polarizations ($>90\%)$ values have been obtained in ferromagnetic films of CrO$_{2}$ by Soulen Jr. and co-workers~\cite{polarization}; polarization values over 85\%~have been reported in ferromagnetic semiconductors based on GaMnAs\cite{polarization2}. Hence, the results above presented are realistic and may be implemented with the state-of-art of experiments.

%

\section{Appendix}

In this section we provide some details used in the calculation of the Green's functions. In particular, we discuss the approximation used to determine the Green's functions along with the determination of the Green's functions equations used to write the corresponding physical quantities presented in the Results and Discussion section.

\subsection{\label{HAapprox}Generalized Mean-Field Approximation}

In deriving the Dyson's equation given by Eq. \eqref{retarded} it is necessary to consider some approximation in order to close the system of equations for the QD Green's function. In fact, the Coulomb correlation at the QD gives rise to an infinite set of equations. Within the Keldysh formalism, both retarded/advanced and ``lesser"~Green's functions are determined as analytic continuations of time-ordered Green's function $\mathbf{G}^{\tau}(\tau,\tau^{\prime})=-i/\hbar\bra \hat{T}_{c}\{\hat{\Psineg}_{d}(\tau)\otimes\hat{\Psineg}^{\dag}_{d}(\tau^{\prime})\}\ket$, where $\hat{T}_{c}$ orders the operators according to their position at the time contour \cite{haug}. The operators are written in the Heisenberg picture whose dynamics is given by the full Hamiltonian $\mathcal{\hat{H}}$, Eq. \eqref{model}. By building the equation of motion for the QD operator, it is possible to determine the equation of motion for $\mathbf{G}^{\tau}(\tau,\tau^{\prime})$. After integrating the contribution from the leads, one ends up with the following expression:
\begin{multline}\label{eq:dyson1}
\mathbf{G}^{\tau}(\tau_{1},\tau_{2})
=
\mathbf{g}^{\tau}(\tau_{1},\tau_{2})
\\+
\int_{c} d\tau_{3}\int_{c} d\tau_{4}~\mathbf{g}^{\tau}(\tau_{1},\tau_{3})\boldsymbol{\Sigma}_{0}(\tau_{3},\tau_{4})\mathbf{G}^{\tau}(\tau_{4},\tau_{2})
\\+
\int_{c} d\tau_{3}~\mathbf{g}^{\tau}(\tau_{1},\tau_{3})\mathbf{U}\mathbf{G}^{\tau(2)}(\tau_{3},\tau_{2})
\end{multline}
where $\boldsymbol{\Sigma}_{0}$ carries the information about the coupling to the leads whose analytic continuation to real time axis gives the retarded/advanced self-energies $\boldsymbol{\Sigma}_{0}^{r/a}$  and the ``lesser'' self-energy $\boldsymbol{\Sigma}^{<}_{0}$ whose expression will be considered in next section. Notice that the last term is a result of the interaction within the QD where the matrix $\mathbf{U}$
\begin{align}\label{intU}
\mathbf{U}=
\begin{pmatrix}
\begin{array}{cccc}
  \mathcal{U} & 0 & 0 & 0 \\
  0 & -\mathcal{U} & 0 & 0 \\
  0 & 0 & \mathcal{U} & 0 \\
  0 & 0 & 0 & -\mathcal{U}
\end{array}
\end{pmatrix}
\end{align}
gives the strength of the interaction and $\mathbf{G}^{\tau(2)}(\tau_{3},\tau_{2})$ is a second order Green'as function containing four operators of the QD. The equation of motion for this Green's function would result in a new equation involving a third order Green's function and so forth. In this way, it is necessary to consider some approximation to truncate the infinite set of equations generated by this technique. To perform such an approximation, we start from the expression for $\mathbf{G}^{\tau(2)}(\tau_{3},\tau_{2})$,
\begin{multline*}
\mathbf{G}^{t(2)}(\tau_{3},\tau_{2})=\\
\begin{pmatrix}
  \bra\bra\hat{d}_{\uparrow}\hat{n}_{d\downarrow}\hat{d}^{\dag}_{\uparrow}\ket\ket^{c}                                                                  &       \bra\bra\hat{d}_{\uparrow}\hat{n}_{d\downarrow}\hat{d}_{\downarrow}\ket\ket^{c}               &        \bra\bra\hat{d}_{\uparrow}\hat{n}_{d\downarrow}\hat{d}^{\dag}_{\downarrow}\ket\ket^{c}                       &  \bra\bra\hat{d}_{\uparrow}\hat{n}_{d\downarrow}\hat{d}_{\uparrow}\ket\ket^{c}
  \\
  \bra\bra\hat{d}^{\dag}_{\downarrow}\hat{n}_{d\uparrow}\hat{d}^{\dag}_{\uparrow}\ket\ket^{c}                                                           &       \bra\bra\hat{d}^{\dag}_{\downarrow}\hat{n}_{d\uparrow}\hat{d}_{\downarrow}\ket\ket^{c}        &         \bra\bra\hat{d}^{\dag}_{\downarrow}\hat{n}_{d\uparrow}\hat{d}^{\dag}_{\downarrow}\ket\ket^{c}               & \bra\bra\hat{d}^{\dag}_{\downarrow}\hat{n}_{d\uparrow}\hat{d}_{\uparrow}\ket\ket^{c}
  \\
  \bra\bra\hat{d}_{\downarrow}\hat{n}_{d\uparrow}\hat{d}^{\dag}_{\uparrow}\ket\ket^{c}                                                                  &       \bra\bra\hat{d}_{\downarrow}\hat{n}_{d\uparrow}\hat{d}_{\downarrow}\ket\ket^{c}               &         \bra\bra\hat{d}_{\downarrow}\hat{n}_{d\uparrow}\hat{d}^{\dag}_{\downarrow}\ket\ket^{c}                      &  \bra\bra\hat{d}_{\downarrow}\hat{n}_{d\uparrow}\hat{d}_{\uparrow}\ket\ket^{c}
  \\
  \bra\bra\hat{d}^{\dag}_{\uparrow}\hat{n}_{d\downarrow}\hat{d}^{\dag}_{\uparrow}\ket\ket^{c}                                                           &       \bra\bra\hat{d}^{\dag}_{\uparrow}\hat{n}_{d\downarrow}\hat{d}_{\downarrow}\ket\ket^{c}        &         \bra\bra\hat{d}^{\dag}_{\uparrow}\hat{n}_{d\downarrow}\hat{d}^{\dag}_{\downarrow}\ket\ket^{c}               & \bra\bra\hat{d}^{\dag}_{\uparrow}\hat{n}_{d\downarrow}\hat{d}_{\uparrow}\ket\ket^{c}
\end{pmatrix}
\end{multline*}
where for compactness we have used the Zubarev notation\cite{Zubarev} in which $\bra\bra\hat{A}\hat{B}\hat{C}\ket\ket^{c}=-i/h\bra\hat{T}_{c}\{\hat{A}(\tau_{3})\hat{B}(\tau_{3})\hat{C}(\tau_{2})\}\ket
$. Notice that $\hat{n}_{d\sigma}=\hat{d}^{\dag}_{\sigma}\hat{d}_{\sigma}$ is the QD number operator for spin $\sigma$.

In order to close the system of equations, we use the following decoupling scheme\cite{franssonbook}:
\begin{multline*}
\bra\bra\hat{d}_{\sigma_{1}}\hat{n}_{\bar{\sigma_{1}}}\hat{d}^{\dag}_{\sigma_{2}}\ket\ket^{c}
\sim
\bra\hat{d}_{\sigma_{1}}\hat{d}^{\dagger}_{\bar{\sigma_{1}}}\ket\bra\bra\hat{d}_{\bar{\sigma_{1}}}\hat{d}^{\dag}_{\sigma_{2}}\ket\ket^{c}
\\-
\bra\hat{d}_{\sigma_{1}}\hat{d}_{\bar{\sigma_{1}}}\ket\bra\bra\hat{d}^{\dagger}_{\bar{\sigma_{1}}}\hat{d}^{\dag}_{\sigma_{2}}\ket\ket^{c}
+
\bra\hat{d}^{\dagger}_{\bar{\sigma_{1}}}\hat{d}_{\bar{\sigma_{1}}}\ket\bra\bra\hat{d}_{\sigma_{1}}\hat{d}^{\dag}_{\sigma_{2}}\ket\ket^{c}
\end{multline*}

\begin{multline*}
\bra\bra\hat{d}_{\sigma_{1}}\hat{n}_{\bar{\sigma_{1}}}\hat{d}_{\sigma_{2}}\ket\ket^{c}
\sim
\bra\hat{d}_{\sigma_{1}}\hat{d}^{\dagger}_{\bar{\sigma_{1}}}\ket\bra\bra\hat{d}_{\bar{\sigma_{1}}}\hat{d}_{\sigma_{2}}\ket\ket^{c}
\\-
\bra\hat{d}_{\sigma_{1}}\hat{d}_{\bar{\sigma_{1}}}\ket\bra\bra\hat{d}^{\dagger}_{\bar{\sigma_{1}}}\hat{d}_{\sigma_{2}}\ket\ket^{c}
+
\bra\hat{d}^{\dagger}_{\bar{\sigma_{1}}}\hat{d}_{\bar{\sigma_{1}}}\ket\bra\bra\hat{d}_{\sigma_{1}}\hat{d}_{\sigma_{2}}\ket\ket^{c}
\end{multline*}
with a similar decoupling for  $\bra\bra\hat{d}^{\dag}_{\sigma_{1}}\hat{n}_{\bar{\sigma_{1}}}\hat{d}_{\sigma_{2}}\ket\ket^{c}$ and $\bra\bra\hat{d}^{\dag}_{\sigma_{1}}\hat{n}_{\bar{\sigma_{1}}}\hat{d}^{\dag}_{\sigma_{2}}\ket\ket^{c}
$. In this scheme, it appears three types of averages each one related with a specific feature of the correlation within the QD. In fact, averages of the form $\bra\hat{d}^{\dagger}_{\sigma_{1}}\hat{d}_{\sigma_{1}}\ket$ represent the Coulomb correlations due to the electron-electron interaction; $\bra\hat{d}_{\sigma_{1}}\hat{d}_{\sigma_{1}}\ket$ and its adjoint are anomalous averages being different of zero due to proximity effect arising from the coupling with to the superconductor. It represents the amplitude of finding a superconductor excitation within the QD. Finally, averages involving $\bra\hat{d}^{\dagger}_{\sigma_{1}}\hat{d}_{\bar{\sigma_{1}}}\ket$ account for spin-flip scattering within the QD. This average is also non-zero for intermediate values of the angle $\theta$ between the magnetization vectors of $F_{1}$ and $F_{2}$. In this way, the electron within the QD can flip its spin as a result of the interplay between the electronic correlation and the misalignment of the magnetization of the ferromagnetic leads.

By substituting the decoupling approximation back into $\mathbf{G}^{t(2)}$ it is possible to write
\begin{align}\label{decsndorder}
\mathbf{G}^{t(2)}(\tau_{3},\tau_{2})\sim\boldsymbol{\Theta}(\tau_{2})\mathbf{G}^{t}(\tau_{3},\tau_{2})
\end{align}
in which the matrix $\boldsymbol{\Theta}$ is written as
\begin{align}\label{Theta}
\boldsymbol{\Theta}=\mathcal{U}
\begin{pmatrix}
  \bra\hat{d}^{\dag}_{\downarrow}\hat{d}_{\downarrow}\ket           &           -\bra\hat{d}_{\uparrow}\hat{d}_{\downarrow}\ket               & \bra\hat{d}_{\uparrow}\hat{d}^{\dag}_{\downarrow}\ket               &                       0
\\
  -\bra\hat{d}^{\dag}_{\downarrow}\hat{d}^{\dag}_{\uparrow}\ket      &          -\bra\hat{d}^{\dag}_{\uparrow}\hat{d}_{\uparrow}\ket           &                       0                                                             & \bra\hat{d}^{\dag}_{\downarrow}\hat{d}_{\uparrow}\ket
\\
  \bra\hat{d}_{\downarrow}\hat{d}^{\dag}_{\uparrow}\ket             &                       0                                                                 & \bra\hat{d}^{\dag}_{\uparrow}\hat{d}_{\uparrow}\ket                 & -\bra\hat{d}_{\downarrow}\hat{d}_{\uparrow}\ket
\\
  0                                                                                 & \bra\hat{d}^{\dag}_{\uparrow}\hat{d}_{\downarrow}\ket                  & -\bra\hat{d}^{\dag}_{\uparrow}\hat{d}^{\dag}_{\downarrow}\ket        &  -\bra\hat{d}^{\dag}_{\downarrow}\hat{d}_{\downarrow}\ket
\end{pmatrix}
\end{align}
whose matrix elements are averages to be determined in self-consistent way for each value of the external parameters. In this work we are interested in the stationary regime in which these averages are taken as time-independent quantities.

With the approximation above it is possible to close the set of equations in order to obtain the Dyson's equation in the time-ordered contour:
\begin{multline}\label{eq:dyson2}
\mathbf{G}^{\tau}(\tau_{1},\tau_{2})
=
\mathbf{g}^{\tau}(\tau_{1},\tau_{2})
\\+
\int_{c} d\tau_{3}\int_{c} d\tau_{4}~\mathbf{g}^{\tau}(\tau_{1},\tau_{3})\boldsymbol{\Sigma}_{0}(\tau_{3},\tau_{4})\mathbf{G}^{\tau}(\tau_{4},\tau_{2})
\\+
\int_{c} d\tau_{3}~\mathbf{g}^{\tau}(\tau_{1},\tau_{3})\boldsymbol{\Theta}(\tau_{3})\mathbf{G}^{\tau}(\tau_{3},\tau_{2}).
\end{multline}

By analytic continuation of Eq. \eqref{eq:dyson2} the relevant Green's functions were determined. The assumption of stationary state allows us to work with Fourier transform of these Green's functions.

\subsection{Self-Consistent Equations}

The approximation we have used leads to the calculation of the averages appearing in Eq. \eqref{Theta}. In order to determine these average values, we use the Keldysh equation obtained by analytic continuation of Eq. \eqref{eq:dyson2}. We have
\begin{align}\label{lesser:final:def2}
\mathbf{G}^{<}(\varepsilon)
=
\mathbf{G}^{r}(\varepsilon)\boldsymbol{\Sigma}^{<}_{0}(\varepsilon)\mathbf{G}^{a}(\varepsilon)
\end{align}
where
\begin{align*}
\boldsymbol{\Sigma}^{<}_{0}(\varepsilon)=\boldsymbol{\Sigma}^{<}_{1}(\varepsilon)+\boldsymbol{\Sigma}^{<}_{2}(\varepsilon)+\boldsymbol{\Sigma}^{<}_{s}(\varepsilon)
\end{align*}
such that each self-energy is determined by using the fluctuation-dissipation theorem since the leads are considered to be in equilibrium. Thus, it is valid to write $\boldsymbol{\Sigma}^{<}_{i}=\mathbf{F}_{i}(\boldsymbol{\Sigma}^{a}_{i}-\boldsymbol{\Sigma}^{r}_{i})$ where we have defined the Fermi matrix,
\begin{align}\label{Fermimatrix}
\mathbf{F}_{i}(\varepsilon)=
\begin{pmatrix}
f_{i} &  0 & 0 &  0 \\
0 &  \bar{f_{i}} & 0 &  0 \\
0 &  0 & f_{i} &  0 \\
0 &  0 & 0 &  \bar{f_{i}}\\
\end{pmatrix},\qquad i=1,2,s
\end{align}
with $f_{i}=f(\varepsilon-eV_{i})$ being the electron Fermi function and $\bar{f_{i}}=f(\varepsilon+eV_{i})$ is the corresponding hole Fermi distribution. Once the superconductor is grounded, then $f_{s}=f(\varepsilon)$ which implies that $\mathbf{F}_{s}$ is diagonal. Considering that $\boldsymbol{\Sigma}^{a}_{i}=[\boldsymbol{\Sigma}^{r}_{i}]^{\dag}$ one can write:
\begin{align*}
\mathbf{\Sigma}_{s}^{<}(\varepsilon)=if(\varepsilon)\Gamma_{s}\widetilde{\varrho}(\varepsilon)
\begin{pmatrix}
1                       &   -\Delta/\varepsilon  &           0                   &      0 \\
-\Delta/\varepsilon  &            1              &           0                   &      0 \\
        0               &            0              &           1                   &\Delta/\varepsilon  \\
        0               &            0              &       \Delta/\varepsilon   &       1
\end{pmatrix}
\end{align*}
where $\widetilde{\varrho}(\varepsilon)=\text{Re}[\varrho(\varepsilon)]$ which is the conventional BCS density of states being different of zero only for $|\varepsilon|>\Delta$.

The contribution from the ferromagnets are given by:
\begin{align}\label{sigma1}
\mathbf{\Sigma}_{1}^{<}(\varepsilon)=i
\begin{pmatrix}
f_{1}\Gamma_{1\uparrow} &  0 & 0 &  0 \\
0 &  \bar{f_{1}}\Gamma_{1\downarrow} & 0 &  0 \\
0 &  0 & f_{1}\Gamma_{1\downarrow} &  0 \\
0 &  0 & 0 &  \bar{f_{1}}\Gamma_{1\uparrow} \\
\end{pmatrix}
\end{align}
for $F_{1}$ and for $F_{2}$ one obtains:
\begin{align}\label{gama2:definiton}
\mathbf{\Sigma}^{<}_{2}(\varepsilon)=i
\begin{pmatrix}
A_{\uparrow}f_{2} & 0 & Bf_{2} & 0 \\
 0 & A_{\downarrow}\bar{f_{2}} & 0 & B\bar{f_{2}} \\
 Bf_{2} & 0 & A_{\downarrow}f_{2} & 0 \\
 0 & B\bar{f_{2}} & 0 & A_{\uparrow}\bar{f_{2}}
\end{pmatrix},
\end{align}
with $A_{\sigma}$ and $B$ being already defined in Eq. \eqref{sigmaF}.

The ``lesser''~Green's function in Nambu space is given by:
\begin{multline*}
\mathbf{G}^{<}(t,t^{\prime})=\dfrac{i}{\hbar}\\
\begin{pmatrix}
\bra\hat{d}^{\dagger}_{\uparrow}(t)\hat{d}_{\uparrow}(t^{\prime})\ket & \bra\hat{d}_{\downarrow}(t)\hat{d}_{\uparrow}(t^{\prime})\ket & \bra\hat{d}^{\dagger}_{\downarrow}(t)\hat{d}_{\uparrow}(t^{\prime})\ket & 0
\\
\bra\hat{d}^{\dagger}_{\uparrow}(t)\hat{d}^{\dagger}_{\downarrow}(t^{\prime})\ket  & \bra\hat{d}_{\downarrow}(t)\hat{d}^{\dagger}_{\downarrow}(t^{\prime})\ket  & 0  & \bra\hat{d}_{\uparrow}(t)\hat{d}^{\dagger}_{\downarrow}(t^{\prime})\ket
\\
\bra\hat{d}^{\dagger}_{\uparrow}(t)\hat{d}_{\downarrow}(t^{\prime})\ket & 0 & \bra\hat{d}^{\dagger}_{\downarrow}(t)\hat{d}_{\downarrow}(t^{\prime})\ket & \bra\hat{d}_{\uparrow}(t)\hat{d}_{\downarrow}(t^{\prime})\ket
\\
0   & \bra\hat{d}_{\downarrow}(t)\hat{d}^{\dagger}_{\uparrow}(t^{\prime})\ket   & \bra\hat{d}^{\dagger}_{\downarrow}(t)\hat{d}^{\dagger}_{\uparrow}(t^{\prime})\ket   & \bra\hat{d}_{\uparrow}(t)\hat{d}^{\dagger}_{\uparrow}(t^{\prime})\ket
\end{pmatrix}
\end{multline*}
and the averages we are looking for in Eq. \eqref{Theta} are obtained by setting $t=t^{\prime}$ in $\mathbf{G}^{<}(t,t^{\prime})$ and performing the Fourier transform. In this case, for instance, the average of any two operators $\bra \hat{d}_{\alpha}\hat{d}_{\beta}\ket$ may be written as
\begin{align*}
\bra \hat{d}_{\alpha}\hat{d}_{\beta}\ket
=
\dfrac{1}{2\pi i}\int\mathbf{G}_{\alpha\beta}^{<}(\varepsilon, \bra \hat{d}_{\alpha}\hat{d}_{\beta}\ket,\cdots)~d\varepsilon
\end{align*}
with $\mathbf{G}_{\alpha\beta}^{<}$ being a matrix element of the Keldysh equation [Eq. \eqref{lesser:final:def2}]. Other matrix elements of $\mathbf{G}^{<}$ correspond to averages of $\bra\hat{d}^{\dag}_{\alpha}\hat{d}_{\beta}\ket$ and $\bra\hat{d}^{\dag}_{\alpha}\hat{d}^{\dag}_{\beta}\ket$. Notice that the $\hbar$ factor was canceled by the one appearing in the Fourier transform. All matrix elements appearing in Eq. \eqref{Theta} may be identified by comparing with the ``lesser'' Green's function matrix.

\bibliography{biblio}

\end{document}